\documentclass[pdflatex,sn-mathphys-num]{sn-jnl}

\usepackage{graphicx}%
\usepackage{multirow}%
\usepackage{amsmath,amssymb,amsfonts}%
\usepackage{amsthm}%
\usepackage{mathrsfs}%
\usepackage[title]{appendix}%
\usepackage{xcolor}%
\usepackage{textcomp}%
\usepackage{manyfoot}%
\usepackage{booktabs}%
\usepackage{algorithm}%
\usepackage{algorithmicx}%
\usepackage{algpseudocode}%
\usepackage{listings}%
\usepackage[inkscapelatex=false]{svg}
\usepackage{upgreek}
\usepackage{xr}
\theoremstyle{thmstyleone}%
\theoremstyle{thmstyletwo}%

\theoremstyle{thmstylethree}%

\begin{document}

\title[Article Title]{Vacuum-dressed superconductivity in NbN observed in a high-$Q$ terahertz cavity}

\author[1]{\fnm{Hongjing} \sur{Xu}}
\author*[2,3,4]{\fnm{Andrey} \sur{Baydin}}\email{baydin@rice.edu}
\author[5]{\fnm{Qinyan} \sur{Yi}}
\author[5]{\fnm{I-Te} \sur{Lu}}
\author[1,2,6]{\fnm{Ningxu} \sur{Zhu}}
\author[2,6]{\fnm{T. Elijah} \sur{Kritzell}}
\author[2,6]{\fnm{Jacques} \sur{Doumani}}
\author[2,6]{\fnm{Dasom} \sur{Kim}}
\author[7]{\fnm{Fuyang} \sur{Tay}}
\author[5,8]{\fnm{Angel} \sur{Rubio}}
\author*[1,2,3,4,9]{\fnm{Junichiro} \sur{Kono}}\email{kono@rice.edu}
\affil[1]{\orgdiv{Department of Physics and Astronomy}, \orgname{Rice University}, \orgaddress{\street{6100 Main Street}, \city{Houston}, \state{TX}, \country{USA}}}
\affil[2]{\orgdiv{Department of Electrical and Computer Engineering}, \orgname{Rice University}, \orgaddress{\street{6100 Main Street}, \city{Houston}, \state{TX}, \country{USA}}}
\affil[3]{\orgdiv{Smalley-Curl Institute}, \orgname{Rice University}, \orgaddress{\street{6100 Main Street}, \city{Houston}, \state{TX}, \country{USA}}}
\affil[4]{\orgdiv{Rice Advanced Materials Institute}, \orgname{Rice University}, \orgaddress{\street{6100 Main Street}, \city{Houston}, \state{TX}, \country{USA}}}
\affil[5]{\orgdiv{Theory Department}, \orgname{Max Planck Institute for the Structure and Dynamics of Matter}, \orgaddress{\city{Hamburg}, \country{Germany}}}
\affil[6]{\orgdiv{Applied Physics Graduate Program, Smalley-Curl Institute}, \orgname{Rice University}, \orgaddress{\city{Houston}, \state{TX}, \country{USA}}}
\affil[7]{\orgdiv{Department of Physics}, \orgname{Columbia University}, \orgaddress{\city{New York}, \state{NY}, \country{USA}}}
\affil[8]{\orgdiv{Initiative for Computational Catalysis and Center for Computational Quantum Physics}, \orgname{The Flatiron Institute}, \orgaddress{\city{New York}, \state{NY}, \country{USA}}}
\affil[9]{\orgdiv{Department of Materials Science and NanoEngineering}, \orgname{Rice University}, \orgaddress{\street{6100 Main Street}, \city{Houston}, \postcode{77005}, \state{TX}, \country{USA}}}

\abstract{Emerging theoretical frameworks suggest that physical properties of matter can be altered within an optical cavity by harnessing quantum vacuum electromagnetic fluctuations, even in the total absence of external driving fields. Among the most intriguing predictions is the potential to noninvasively manipulate superconductivity. Here, we experimentally observe modified superconductivity in niobium nitride (NbN) thin films within high-quality-factor ($Q$) terahertz cavities. Using terahertz time-domain spectroscopy, we characterize the NbN response both in free space and within a high-$Q$ photonic-crystal cavity. Our analysis reveals significant cavity-induced modifications to the optical conductivity. A theoretical model indicates that these changes originate from a substantial ($\sim13\,\%$) reduction in the superfluid density and a minor ($\sim2\,\%$) reduction in the superconducting gap, driven by cavity vacuum fluctuations. These results demonstrate a platform for engineering ground states via vacuum--matter coupling, opening frontiers in cavity materials science.
}

\keywords{Cavity quantum electrodynamics, Superconductivity, THz spectroscopy, Cavity materials engineering, Condensed matter physics}

\maketitle

\section{Introduction}\label{sec1}

Light--matter interaction has been explored as a powerful means to manipulate the properties of superconductors since the 1960s, when researchers demonstrated that the critical current of thin superconducting films could be enhanced by subgap microwave radiation---a phenomenon known as the Dayem--Wyatt effect~\cite{wyatt_microwave-enhanced_1966,dayem_behavior_1967,wyatt_increase_1971}. This effect was later explained within the Eliashberg theory, which posits that the presence of microwave electromagnetic fields causes a nonequilibrium but stationary redistribution of quasiparticles, leading to a modification of the superconducting gap~\cite{eliashberg_film_1970,ivlev_influence_1971,eliashberg_inelastic_1972,ivlev_nonequilibrium_1973,klapwijk_radiation-stimulated_1977}. As laser technology advanced, it became possible to use strong, ultrashort infrared pulses to resonantly drive low-energy excitations, such as phonons, to induce, via nonlinear phononics, a superconducting-like optical response at temperatures well above the equilibrium critical temperature, $T_\mathrm{c}$~\cite{mitrano_possible_2016,cavalleri_photo-induced_2018,rowe_resonant_2023,bloch_strongly_2022,disa_engineering_2021}. 

In recent years, numerous theoretical proposals have emerged to control or enhance superconductivity using light--matter hybrid states inside a cavity. One proposed mechanism is the quantum Eliashberg effect, wherein the classical electromagnetic driving field is replaced by fluctuating quantum-vacuum fields to enhance the $T_\mathrm{c}$ of a BCS superconductor~\cite{curtis_cavity_2019}. A different approach predicts a $T_\text{c}$ enhancement in the phonon-mediated superconductor MgB$_2$ by approximately 10\% by coupling the cavity's vacuum field with electrons~\cite{lu_cavity-enhanced_2024,lu_cavity_2025}. 
Furthermore, other earlier research has examined the formation of cavity phonon-polaritons and their capacity to modify electron--phonon coupling, thereby modulating the $T_\mathrm{c}$ of a monolayer of FeSe deposited on $\mathrm{SrTiO_3}$~\cite{sentef_cavity_2018}.  

\begin{figure}[h]
\centering
\includegraphics[width=0.9\textwidth]{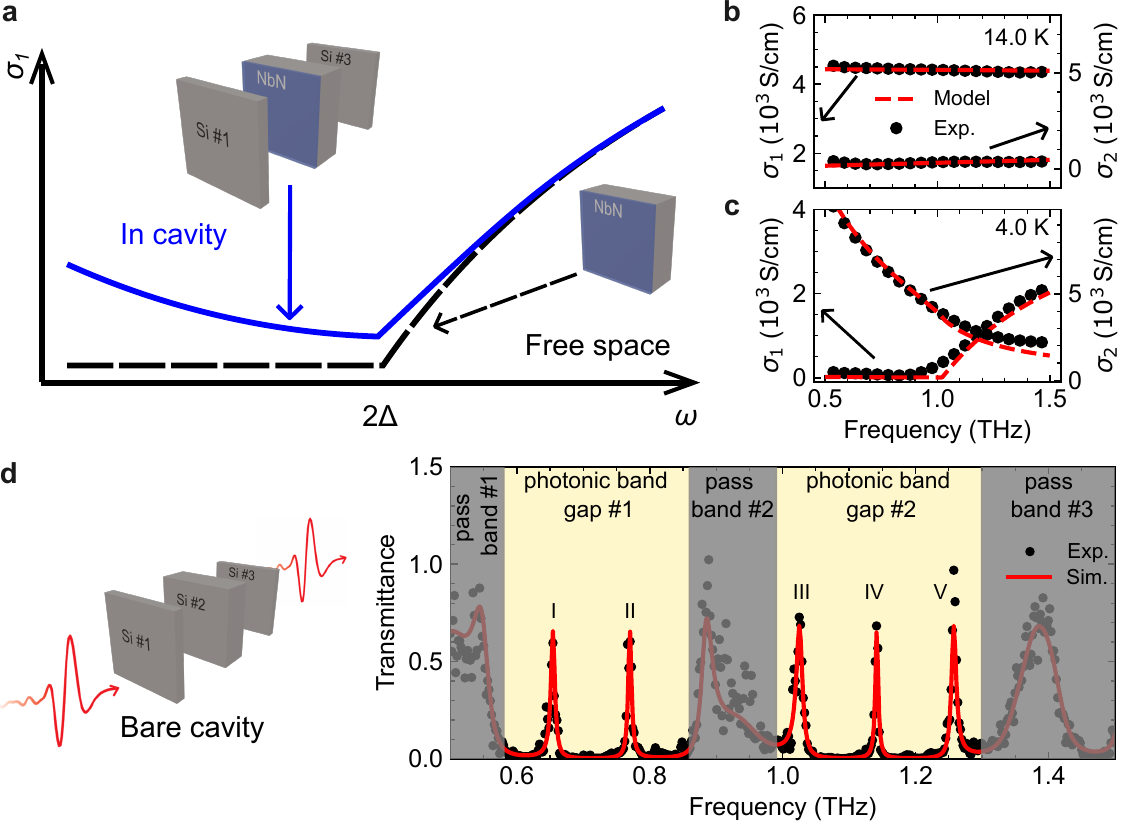}
\caption{\textbf{Probing cavity-modified superconductivity in NbN.} (\textbf{a})~Conceptual illustration of the vacuum--matter coupling-induced change in the real part of the optical conductivity, $\sigma_1(\omega)$, of a NbN film inside a high-$Q$ one-dimensional photonic-crystal cavity (1D-PCC). In free space, $\sigma_1$ is zero below the superconducting optical gap $2\Delta$. Within the 1D-PCC, the confined quantum vacuum fluctuations ``dress'' the superconducting ground state, making $\sigma_1$ finite below the gap. (\textbf{b})~Optical conductivity $\sigma = \sigma_1 + i\sigma_2$ of a 10-nm NbN film measured in free space at $14$\,K in the normal state. (\textbf{c})~$\sigma$ of the NbN film in free space at $4$\,K in the superconducting state. The real part shows the opening of the superconducting gap, while the imaginary part exhibits the $1/\omega$ inductive response characteristic of a superfluid. Symbols represent experimental data and solid lines represent fits using the Zimmermann model. (\textbf{d})~Transmittance of the bare 1D-PCC at $6\,\mathrm{K}$. Experimental data (black dots) are in excellent agreement with the transfer-matrix method simulation (red line). Gold shaded regions indicate the photonic band gaps containing high-$Q$ defect modes (peaks I, II, III, IV, and V), while gray regions represent the pass bands. The thicknesses of the silicon layers \#1, \#2, and \#3 are $94$, $348$, and $97$\,$\upmu$m, respectively. The spacing between silicon layers \#1 and \#2 (\#2 and \#3) is $77$\,$\upmu$m ($76$\,$\upmu$m).} \label{Intro_Fig}
\end{figure}

Despite numerous theoretical predictions, experimental evidence of cavity-altered superconductivity has remained scarce~\cite{thomas_exploring_2025,keren_cavity-altered_2025}. While a few reports have explored cavity materials engineering in other contexts~\cite{paravicini-bagliani_magneto-transport_2019,enkner_tunable_2025,jarc_cavity-mediated_2023,baydin_perspective_2025}, a direct observation of modified superconductivity remains a challenge. In Ref.~\cite{keren_cavity-altered_2025}, Keren \textit{et al}.\ demonstrated that coupling $\mathrm{\kappa}$-$\mathrm{(BEDT}$-$\mathrm{TTF)_2}$$\mathrm{Cu\left[N\left(CN\right)_2\right]Br}$ 
to the evanescent, confined hyperbolic modes of hBN leads to resonant cavity--matter interactions that suppress the superfluid density at the interface. These results indicate that the electromagnetic vacuum in a dark cavity can directly modify the superconducting ground state. Here, we investigate the optical properties of a film of the BCS superconductor niobium nitride (NbN) embedded within a high-quality-factor ($Q$) terahertz (THz) one-dimensional photonic-crystal cavity (1D-PCC), comparing the response to its free-space counterpart at cryogenic temperatures; see Fig.~\ref{Intro_Fig}a. We observe a significant discrepancy between the measured transmittance of the cavity--NbN system and calculations based on the measured free-space complex optical conductivity of the NbN film. By matching the lineshapes of the cavity modes, we extract the effective complex optical conductivity ``dressed'' by quantum vacuum fluctuations. Crucially, the real part of the optical conductivity exhibits a substantial increase around the superconducting gap energy below $9$\,K. Our results provide direct evidence that the fundamental electrodynamics of NbN superconducting films are intrinsically altered by quantum vacuum fluctuations that are enhanced in the THz cavity. \\

\section{Results}\label{sec2}

\noindent\textbf{Superconducting electrodynamics of NbN in free space} 

\noindent We measured the complex optical conductivity, $\sigma(\omega) = \sigma_1(\omega) + i \sigma_2(\omega)$ where $\omega$ is the frequency of the weak external probing light, of a 10-nm-thick NbN film from $4$ to $14$\,K using THz time-domain spectroscopy (THz-TDS). The conductivity was extracted using the Tinkham formula~\cite{glover_conductivity_1957}; see Materials and Methods in the Supplementary Information (SI). The spectral bandwidth for analysis extended between $0.5$ and $1.5$\,THz. Direct current (DC) electrical transport measurements established the superconducting transition temperature of the NbN film on its silicon substrate at $T_{\mathrm{c}} = 13.1$\,K; see SI. 

Figure~\ref{Intro_Fig}b and c show representative optical conductivity spectra; see SI for the complete temperature dependence. In the normal state ($T>13.1\,\mathrm{K}$), $\sigma_{1}(\omega)$ is frequency-independent, indicating a large Drude scattering rate, while $\sigma_{2}(\omega)$ remains near zero (Fig.~\ref{Intro_Fig}b). A Drude fit at $14$\,K yields a DC conductivity of $\sigma_{\mathrm{DC}}=4.44 \times 10^{2} \,\mathrm{S/cm}$. In the superconducting state, $\sigma_{1}$ exhibits a temperature-dependent gap, and $\sigma_{2}$ displays the characteristic 1/$\omega$ dependence associated with the superfluid response (Fig.~\ref{Intro_Fig}c). The gap in $\sigma_{1}$ closes as the temperature approaches $T_\mathrm{c}$. We modeled the superconducting-state conductivity using the Zimmermann formula~\cite{zimmermann_optical_1991}, utilizing the normal-state conductivity $\sigma_{\mathrm{n}}$ measured at $14\,\mathrm{K}$ to extract the superconducting optical gap, 2$\Delta$. The model shows excellent agreement with the experimental data; see SI. The extracted gap for the free-space NbN film is $2\Delta(4\,\mathrm{K}) = 1.02\,\mathrm{THz} = 4.21\,$meV. Given $k_\text{B} T_{\mathrm{c}} = 1.12$\,meV with $T_{\mathrm{c}} = 13.1$\,K, the ratio $2\Delta / k_\text{B} T_{\mathrm{c}}$ is $3.76$, which is close to the BCS weak-coupling limit ($3.53$), as expected for the phonon-mediated BCS superconductor NbN. \\

\noindent\textbf{Bare cavity characterization.} 

\noindent Before investigating the interaction of the 10-nm-thick NbN film with quantum-vacuum fields inside a 1D-PCC, we characterized the bare 1D-PCC, which consisted of two high-resistivity silicon plates with thicknesses of $94$ and $97$\,$\upmu$m, two vacuum layers ($77$ and $76$\,$\upmu$m), and a central defect layer. The defect layer is a silicon substrate with a thickness of $348$\,$\upmu$m that breaks the translational symmetry of the layered structure (Fig.~\ref{Intro_Fig}d). All layer thicknesses were verified using a micrometer. The silicon plates and the vacuum layers act as Bragg mirrors, creating photonic band gaps (regions of vanishing transmission) in the THz regime. The symmetry-breaking induced by the thick central substrate gives rise to sharp transmittance peaks, corresponding to high-$Q$ cavity defect modes, inside the photonic band gaps.

This cavity was designed so that five cavity defect modes reside near the superconducting optical gap of NbN, 2$\Delta=1.02\,\mathrm{THz}$ at very low temperatures. As $2\Delta$ decreases with increasing temperature, the gap frequency effectively ``sweeps through'' different cavity defect modes. The amplitude and frequency of each mode are highly sensitive to the complex optical conductivity of the NbN thin film, owing to the high $Q$ factors. The high $Q$ values of the cavity modes—reaching the order of $100$—are crucial to this study, as they provide the spectral resolution necessary to resolve small, cavity-induced changes in the optical conductivity that would otherwise be obscured by broader resonance linewidths. Therefore, the transmittance spectra of the NbN-cavity system provide a sensitive probe of the temperature dependence of $\sigma(\omega,T)$. 

We measured the transmittance of the bare 1D-PCC in the same THz-TDS setup at $6\,\mathrm{K}$ (Fig.~\ref{Intro_Fig}d). We used the transfer-matrix method (TMM) to simulate the THz transmittance spectra using the experimentally measured thicknesses and refractive indices of all cavity layers. As shown in Fig.~\ref{Intro_Fig}d, the TMM simulation matches the experimental spectra both in terms of the cavity mode positions and the edges of the photonic band gaps. This confirms the validity of the values used in the TMM, allowing them to be applied in the analysis of the NbN-integrated structure. Furthermore, the TMM calculations show that the electric field is maximized at the surface of the defect layer (SI), where the NbN thin film is located, ensuring strong vacuum--matter coupling.  \\

\begin{figure}[h]
\centering
\includegraphics[width=0.75\textwidth]
{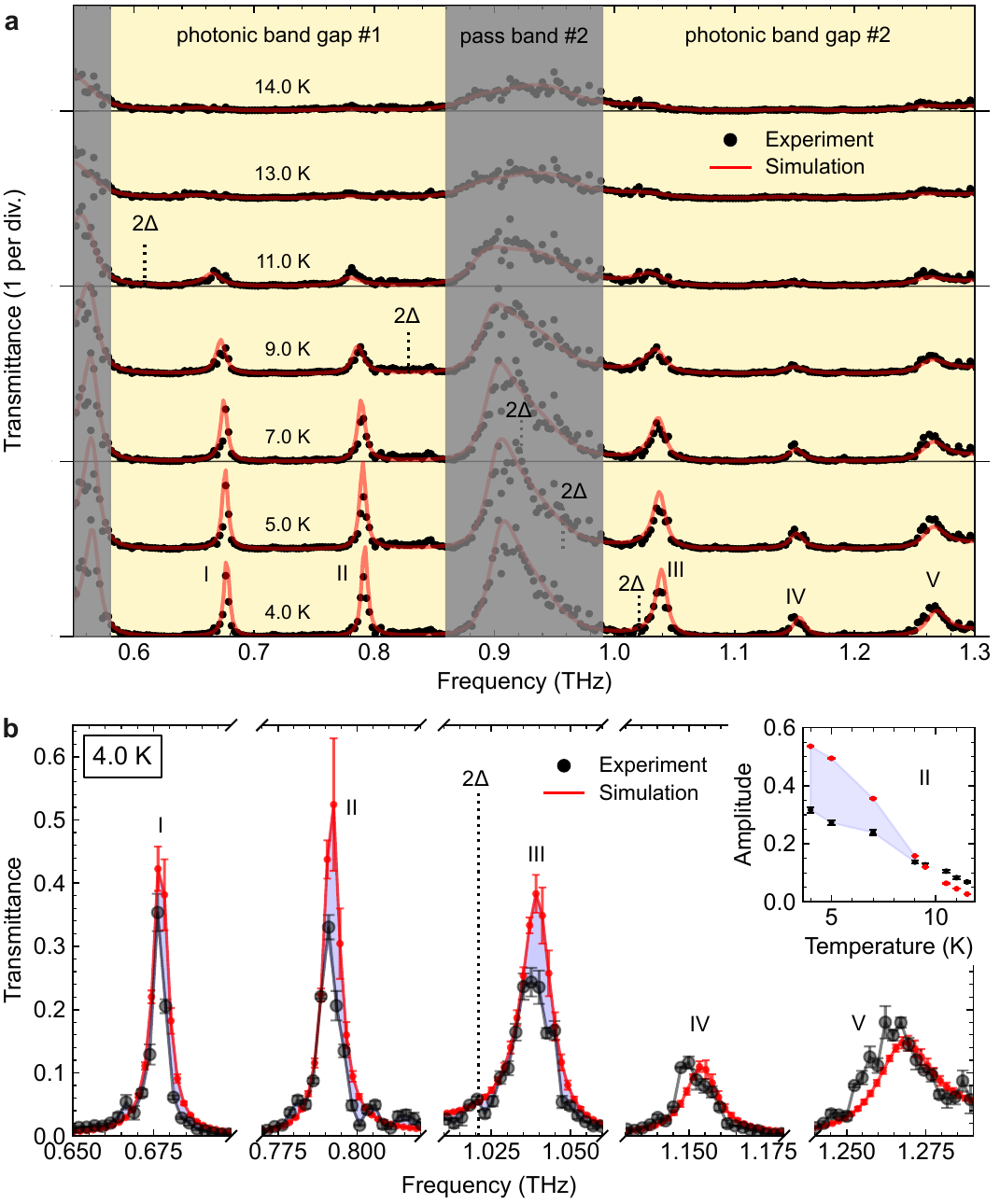}
\caption{\textbf{Temperature dependence of the NbN-cavity transmittance spectra.}
(\textbf{a})~Comparison between simulated (red solid lines) and experimental (black solid dots) transmittance of the NbN-cavity integration at various temperatures from $4.0$ to $14.0$\,K. The transmittance curves are vertically offset for clarity. The vertical dotted lines track the temperature-dependent optical gap, $2\Delta(T)$. The regions of interest -- the photonic band gaps where high-$Q$ defect modes reside -- are highlighted in gold, and the gray regions indicate the pass bands. 
(\textbf{b})~High-resolution experiment-simulation comparison of the cavity modes at $T=4.0\,\mathrm{K}$. The amplitude discrepancies of peaks~I, II, and III (shaded regions) are larger than the combined uncertainties of the experimental measurement and the TMM simulation. The inset shows the temperature dependence of the amplitude for peak~II, illustrating the discrepancy between simulation and experiment that emerges exclusively in the superconducting state.}\label{Cavity_T}
\end{figure}

\noindent\textbf{Cavity-induced changes in the optical conductivity of NbN}

\noindent Figure~\ref{Cavity_T}a shows the transmittance spectra for the NbN-cavity integration at various temperatures from $4$ to $14$\,K. Black dots represent experimental data, and red solid lines represent simulated spectra using the TMM. We determined the complex refractive index of the NbN film from its measured free-space $\sigma(\omega)$ and incorporated it into the TMM to simulate the NbN-cavity transmittance spectra. At $T>13.1\,\mathrm{K}$, the simulated and the experimental spectra coincide across the full frequency range, indicating that the normal-state optical conductivity is not affected by the cavity.

As $T$ decreases, the modes above the optical gap $2\Delta(T)$ (peaks~IV and V) show no significant discrepancy between simulation and experiment. In contrast, modes near and below $2\Delta(T)$ (peaks~I, II, and III) show a considerable decrease in peak amplitude in the experiment compared to the TMM simulation, especially below $9\,\mathrm{K}$. This discrepancy indicates that the optical conductivity of the NbN film is modified when integrated into the cavity. Note that the simulation is constrained by the experimentally obtained refractive index values for all individual components of the cavity. 

The simulated and experimental spectra at $4$\,K, including error bars, are shown in Fig.~\ref{Cavity_T}b; the inset plots the amplitude of peak~II (obtained via Lorentzian fits) as a function of $T$. These results highlight that the discrepancy between simulation and experiment exists only in peaks I, II, and III within the superconducting state; see SI for further details in the peak analysis. This suppression of the cavity resonance suggests that the coupling between the superconducting film and the cavity's confined electromagnetic field induces a redistribution of the oscillator strength not present in free space. \\

\noindent\textbf{Data interpretation and theoretical modeling}

\noindent The primary experimental finding here is that embedding the NbN film in a 1D-PCC enhances its absorption near and below the optical gap $2\Delta$ in the superconducting state. To quantify this effect, we integrated the Zimmermann model into the TMM simulation to fit the NbN-cavity transmittance at $4\,\mathrm{K}$, where the superconducting gap is at its maximum and the divergence from free-space behavior is most pronounced.

A critical factor in this analysis is the high $Q$-factor of our cavity modes; see Table~\ref*{supp-tab:sup_bare_Q} of SI. The extremely narrow linewidths of the resonances provide a high-resolution ``spectral ruler''; that is, even subtle modifications to the complex optical conductivity result in measurable shifts in the peak amplitudes and positions. This sensitivity allowed us to claim precise cavity-induced changes in the complex optical conductivity that would be indistinguishable in a low-finesse system.

The only two adjustable fitting parameters in this procedure were the cavity-modified gap $\Delta_\text{in-cav}$ and the effective temperature $T_\mathrm{eff}$. At each experimental temperature $T$, we optimized these parameters to best fit the transmittance spectrum of the NbN-cavity structure (Fig.~S10 of SI). We emphasize that $T_\mathrm{eff}$ is a phenomenological parameter used to model the increased quasiparticle density; it represents a nonthermal redistribution of the electronic population rather than physical heating of the sample, which was maintained at a bath temperature of $4$\,K. Once the cavity-dressed optical conductivity $\sigma_\text{in-cav}$ was obtained, we calculated the new superfluid density $n_\text{s,in-cav}$ from the inductive response $\sigma_\text{2,in-cav} = (n_\text{s,in-cav}e^2/m^*)\,\omega^{-1}$ in the $\omega \rightarrow 0$ limit.

\begin{figure}[h]
\centering
\includegraphics[width=0.8\textwidth]{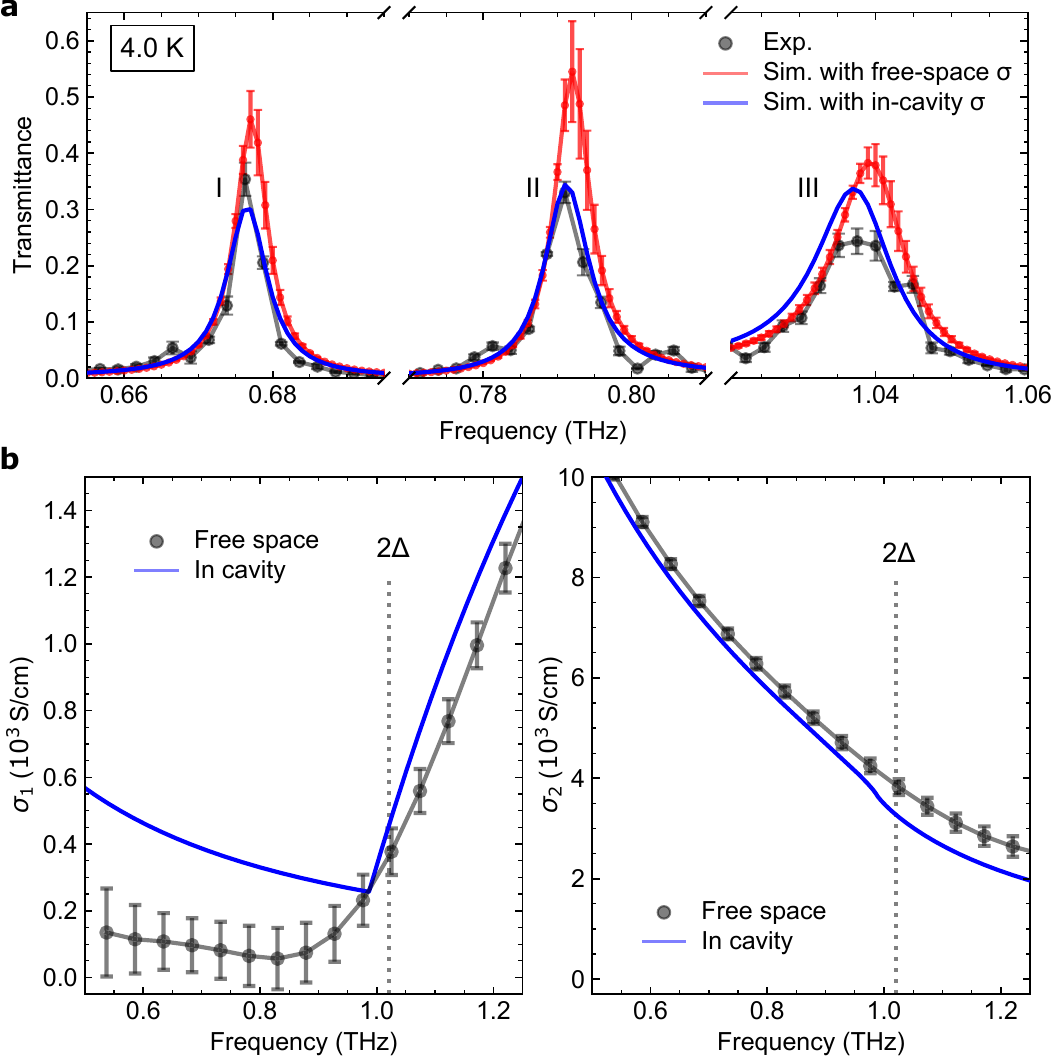}
\caption{\textbf{Modified complex optical conductivity of the NbN film embedded in the 1D-PCC at $4\,\mathrm{K}$.} The extraction is based on matching the cavity mode transmittance via the TMM simulations and the experiment. In the TMM simulations, we used the Zimmermann modeling parameters (effective temperature and superconducting gap) to fit the experimental transmittance spectra. (\textbf{a})~Detailed amplitude comparison of experimental data (black circles), TMM simulations using free-space conductivity (red line), and TMM fits using the cavity-dressed conductivity (blue line) for cavity modes I, II, and III. The high $Q$-factors of these modes provide the sensitivity required to resolve the discrepancy between free-space and in-cavity responses. (\textbf{b})~Extracted real ($\sigma_1$, left) and imaginary ($\sigma_2$, right) parts of the modified complex conductivity of the NbN film. The black circles represent the experimentally measured conductivity of NbN in free space, and the blue solid line represents the cavity-dressed conductivity with the fitted parameters ($T_{\text{eff}}$ and $\Delta_{\text{in-cav}}$). The vertical dotted lines indicate the superconducting optical gap $2\Delta$ of NbN in free space at $4\,\mathrm{K}$.}\label{4K_Extract}
\end{figure}

Here, we focus on the results obtained at $4$\,K, where the observed discrepancy in the transmittance is the largest. Figure~\ref{4K_Extract}a shows that the fitted spectra (blue solid lines) near (peak III) and below the optical gap (peaks I and II) quantitatively reproduce the experimental data. Our optimal fit yielded $\Delta_\text{in-cav}/\Delta(4\,\mathrm{K}) = 0.977\pm0.008$ and $T_\mathrm{eff} = 7.5\pm0.2\,\mathrm{K}$, corresponding to a superfluid density ratio of $n_\text{s,in-cav}/n_{\mathrm{s}}(4\,\mathrm{K}) = 0.87\pm0.03$. The slight overestimation of the amplitude of peak III stems from the intrinsic limitations of the Zimmermann model in describing the conductivity in this specific frequency range, a behavior also observed in free space. The resulting cavity-dressed complex optical conductivity $\sigma_\text{in-cav}(\omega)$ is shown in Fig.~\ref{4K_Extract}e and f (blue solid lines). We observe an increase in $\sigma_1$ and a decrease in $\sigma_2$ below the optical gap frequency. This modification suggests that quantum fluctuations within the high-$Q$ cavity partially break Cooper pairs, thereby suppressing the superfluid density and enhancing optical absorption.

First-principles quantum electrodynamical density functional theory calculations~\cite{lu_cavity-enhanced_2024,lu_cavity_2025,ruggenthaler_quantum-electrodynamical_2014} (see SI for details) predict that, for extremely strong light--matter coupling achievable only in highly subwavelength polaritonic structures~\cite{herzig2024high}, the superconducting transition temperature could increase by up to $0.3$\,K. Based on our experimental observation (SI) that $T_{\mathrm{c}}$ does not change significantly---likely due to a smaller light--matter coupling strength in the 1D-PCC---it is assumed that $T_{\mathrm{c}}$ and the superconducting gap are nearly constant in the calculation (SI). Under this regime, the \textit{ab initio} calculations (restricted to the electron density of the normal phase) show that coupling to the cavity nonetheless modifies the electron density of the NbN film compared with free space, indicating that quantum fluctuations induce a redistribution of electronic charge. This calculated density redistribution is consistent with the minor fractional change in the superconducting gap and the larger suppression of superfluid density observed here.  In this picture, the cavity primarily redistributes electrons and reduces the superfluid density, thereby enhancing absorption, rather than substantially altering the intrinsic pairing gap of NbN, in contrast to $\mathrm{MgB_2}$, where pairing is enhanced and the $T_\mathrm{c}$ is increased~\cite{lu_cavity-enhanced_2024}. \\

\section{Conclusion}\label{sec13}
\noindent We investigated the superconductivity of a NbN thin film integrated into a one-dimensional photonic-crystal cavity using terahertz time-domain spectroscopy. By comparing experimental transmittance data with transfer-matrix method simulations based on the film's free-space refractive index, we observed significant suppressions in cavity mode amplitudes near and below the superconducting optical gap, $2\Delta$. The high $Q$-factor of the dielectric cavity was instrumental in this discovery, providing the narrow spectral resonances necessary to resolve minute modifications in the film's complex optical conductivity. This modification gradually vanishes as the temperature approaches the superconducting transition temperature $T_\mathrm{c}$, and notably occurs in the absence of an external electromagnetic pump. 

Our results suggest that cavity vacuum fluctuations within the ``dark'' cavity modify the optical conductivity by reducing the superfluid density of the embedded NbN film. Analogous to a two-fluid model, the virtual photons of the cavity modes exchange energy with the NbN film, thereby increasing the quasiparticle population at the expense of Cooper pairs. By demonstrating this vacuum-mediated modification of $\sigma(\omega)$ with high precision, our work introduces a robust method to tailor the ground-state properties of BCS superconducting films through cavity engineering and highlights the potential of high-$Q$ architectures for fundamental studies of light--matter interaction phenomena driven by quantum fluctuations in materials.

\section*{Methods}\label{sec11}
\subsection*{NbN film characterization}

\noindent A 10-nm-thick NbN film was deposited on a high-resistivity single-crystal Si (001) substrate heated to 550 °C using a DC magnetron sputtering system~\cite{shi_nbn_2022,zhang_sputtering_2024}. The electrical resistance of a NbN film from the same batch was measured in a Quantum Design Dynacool system, and its onset superconducting transition temperature was $13.1$\,K, consistent with the modeling of the free-space THz conductivity of the NbN film. The resistance--temperature relationship is shown in fig.~\ref*{supp-fig:NbN_free_RT_300} in SI. \\

\subsection*{1D-PCC fabrication}
\noindent The bare 1D-PCC consisted of two Bragg reflector structures, namely a thin high-resistivity intrinsic silicon wafer and a vacuum layer gapped by a spacer (a Kapton film with a 5-mm diameter through-hole in the middle), and a middle defect layer, namely the silicon substrate, which breaks the translational symmetry of the layered structure. The NbN-cavity system had the NbN thin film deposited on the middle silicon substrate. The thickness of each layer was measured with a micrometer; see Fig.~\ref{Intro_Fig}d captions.\\

\subsection*{THz time-domain spectroscopy}
\noindent We performed THz time-domain spectroscopy in an Oxford Instruments Spectromag superconducting magnet system, where the temperature was controlled by balancing the cooling power of flowing liquid helium with a heater installed near the cavity inside the dynamic variable temperature insert (VTI). The whole cavity structure was uniformly cooled down by cold helium vapor. We were able to achieve temperatures below $4.2$\,K by pumping the VTI to reduce the vapor pressure of helium. A linearly polarized THz beam (0.25--2.5\,THz) was generated by optical rectification in a ZnTe crystal, and the transmitted THz signal was measured in another ZnTe crystal by electro-optical sampling. \\

\subsection*{Extraction of the THz complex conductivity of the NbN film in free space}

\noindent We used the Tinkham formula
\begin{align}
\sigma(\omega) = \frac{N_\mathrm{sub}+1}{Z_0 d}\left(\frac{E_\mathrm{sub}(\omega)}{E_\mathrm{film+sub}(\omega)} -1 \right)
\end{align}
to extract the THz complex conductivity of the NbN film from the complex transmission function from the free-space THz-TDS measurements~\cite{glover_conductivity_1957}. Here, $N_\mathrm{sub}$ is the complex refractive index of the silicon substrate, $E_\mathrm{sub}(\omega)/E_\mathrm{film+sub}(\omega)$ is the inverse of the complex electric field transmission of the combined system of the NbN film and the Si substrate (compared to the bare Si substrate), $Z_0 = 376.73\,\Omega$ is the impedance of free space, and $d=10\,$ nm is the thickness of the NbN film. We then calculated the complex relative permittivity $\varepsilon_\mathrm{r}(\omega)$ and refractive index $N(\omega)$ via
\begin{align}
    N(\omega) = \sqrt{\varepsilon_\mathrm{r}(\omega)} = 1 + i\,\frac{{\sigma(\omega)}}{\omega\varepsilon_0},
\end{align}
where $\varepsilon_0$ is the vacuum permittivity. The extracted free-space $\sigma(\omega)$ agrees with the Zimmermann model~\cite{zimmermann_optical_1991}, as detailed in SI.

\backmatter

\bmhead{Supplementary information}
Supplementary Text, Figs. S1 to S10, and Table S1.

\bmhead{Acknowledgements}
The authors thank Motoaki Bamba, Alexander Potts, and Michael Ruggenthaler for useful discussions.

\section*{Declarations}

\begin{itemize}
\item Funding

J.K.\ acknowledges support from the U.S.\ Army Research Office (through Award No.\ W911NF-21-1-0157, W911NF-23-1-0410), the Gordon and Betty Moore Foundation (through Grant No.\ 11520), and the Robert A.\ Welch Foundation (through Grant No.\ C-1509). This work was supported by the European Research Council (ERC-2024-SyG- 101167294; UnMySt), the Cluster of Excellence Advanced Imaging of Matter (AIM), Grupos Consolidados y Alto Rendimiento UPV/EHU, Gobierno Vasco (IT1453-22). We acknowledge support from the Max Planck-New York City Center for Non-Equilibrium Quantum Phenomena. The Flatiron Institute is a division of the Simons Foundation.

\item Conflict of interest/Competing interests 

The authors declare no competing interests
\item Ethics approval and consent to participate

Not applicable
\item Consent for publication

Not applicable
\item Data availability 

The data are available from the authors upon request

\item Code availability 

The codes are available from the authors upon request. 
\item Author contribution: 

H.X., A.B., and J.K.\ conceived and planned the experiments. H.X.\ carried out the THz-TDS experiments. H.X., A.B., N.Z., and I.-T.L.\ analyzed the data. Q.Y.\ and I.-T.L.\ carried out numerical calculations. T.E.K.\ and J.D.\ contributed to sample preparation. D.K.\ and F.T.\ contributed to the experimental setup.  A.B.\ and J.K.\ supervised the project. H.X., A.B., Q.Y., I.-T.L., and J.K.\ prepared the manuscript with inputs from all authors. 
\end{itemize}

\bibliography{NbN_Manuscript_main,SI-bibliography,NbN_SM_1}

\renewcommand{\thefigure}{S\arabic{figure}} 


\begin{center}
    {\huge Supplementary Information}    
\end{center}
\setcounter{section}{0}

\section{Optical conductivity of the NbN thin film in free space}\label{sec1}
\begin{figure}[h]
    \centering
    \includegraphics[width=0.6\linewidth]{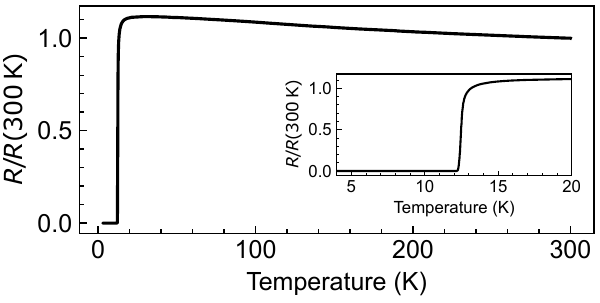}
    \caption{\textbf{The electrical resistance of a NbN film in free space, normalized by the resistance at $300$\,K.} The inset zooms in at the temperature region near $T_\mathrm{c}$. 
	}
    \label{fig:NbN_free_RT_300}
\end{figure}

We measured the low-frequency ($f=9.15\,\mathrm{Hz}$) AC resistance of the NbN film in a Quantum Design Dynacool system, and it could be regarded as the DC resistance due to the $f$ being very small. The electrical resistance--temperature relationship is shown in fig.~\ref{fig:NbN_free_RT_300}. 

We simulated the frequency- and temperature-dependent complex optical conductivity of the superconducting state, $\sigma(\omega,T)=\sigma_{1}(\omega,T)+i\sigma_{2}(\omega,T)$, of the NbN thin film on the silicon substrate using the Zimmermann model~\cite{pracht.heintze.ea_2013,zimmermann.brandt.ea_1991}. 
The Zimmermann formula requires the DC conductivity $\sigma_{\mathrm{DC}}$, the relaxation time $\tau$ (for taking into account impurity scattering), the superconducting transition temperature $T_{\mathrm{c}}$, and the temperature-dependent superconducting gap $\Delta(T)$: 
\begin{equation}\label{eq:zimmermann-formula}
    \sigma(\omega,T) = i\,\frac{\sigma_{\rm{DC}}}{2\omega\tau}\times\left(\int_{\Delta(T)}^{\hbar\omega+\Delta(T)}I_{1}dE+\int_{\Delta(T)}^{\infty}I_{2}dE\right),
\end{equation}
where 
\begin{equation}
\begin{aligned}
I_{1}=\tanh\left(\frac{E}{2k_\mathrm{B}T}\right)\times&\Bigg\{\left[1-\frac{\Delta^{2}(T)+E(E-\hbar\omega)}{P_{3}P_{2}}\right]\frac{1}{P_{3}+P_{2}+i\hbar\tau^{-1}} \\ 
& -\left[1+\frac{\Delta^{2}(T)+E(E-\hbar\omega)}{P_{3}P_{2}}\right]\frac{1}{P_{3}-P_{2}+i\hbar\tau^{-1}}\Bigg\}
\end{aligned}
\end{equation}
and
\begin{equation}
\begin{aligned}
I_{2} & =\tanh\left(\frac{E+\hbar\omega}{2k_\mathrm{B}T}\right)\times&&\Bigg\{\left[1+\frac{\Delta^{2}(T)+E(E+\hbar\omega)}{P_{1}P_{2}}\right]\frac{1}{P_{1}-P_{2}+i\hbar\tau^{-1}} && \\ 
& && -\left[1-\frac{\Delta^{2}(T)+E(E+\hbar\omega)}{P_{1}P_{2}}\right]\frac{1}{-P_{1}-P_{2}+i\hbar\tau^{-1}}\Bigg\} &&\\
& + \tanh\left(\frac{E}{2k_\mathrm{B}T}\right)\times &&\Bigg\{\left[1-\frac{\Delta^{2}(T)+E(E+\hbar\omega)}{P_{1}P_{2}}\right]\frac{1}{P_{1}+P_{2}+i\hbar\tau^{-1}} && \\
& && - \left[1+\frac{\Delta^{2}(T)+E(E+\hbar\omega)}{P_{1}P_{2}}\right]\frac{1}{P_{1}-P_{2}+i\hbar\tau^{-1}}\Bigg\} &&,
\end{aligned}
\end{equation}
with $P_{1}=\sqrt{(E+\hbar\omega)^{2}-\Delta^{2}(T)}$, $P_{2}=\sqrt{E^{2}-\Delta^{2}(T)}$, and $P_{3}=\sqrt{(E-\hbar\omega)^{2}-\Delta^{2}(T)}$. The first integral, $I_1$, is the contribution from the Cooper pairs, while the second integral, $I_2$, is the contribution from the thermal electrons. 
We used and implemented a convenient numerical version of the above formula developed in Ref.~\cite{zimmermann.brandt.ea_1991} in our in-house Python code. 

\begin{figure}[htbp]
    \centering
    \includegraphics[width=0.68\linewidth]{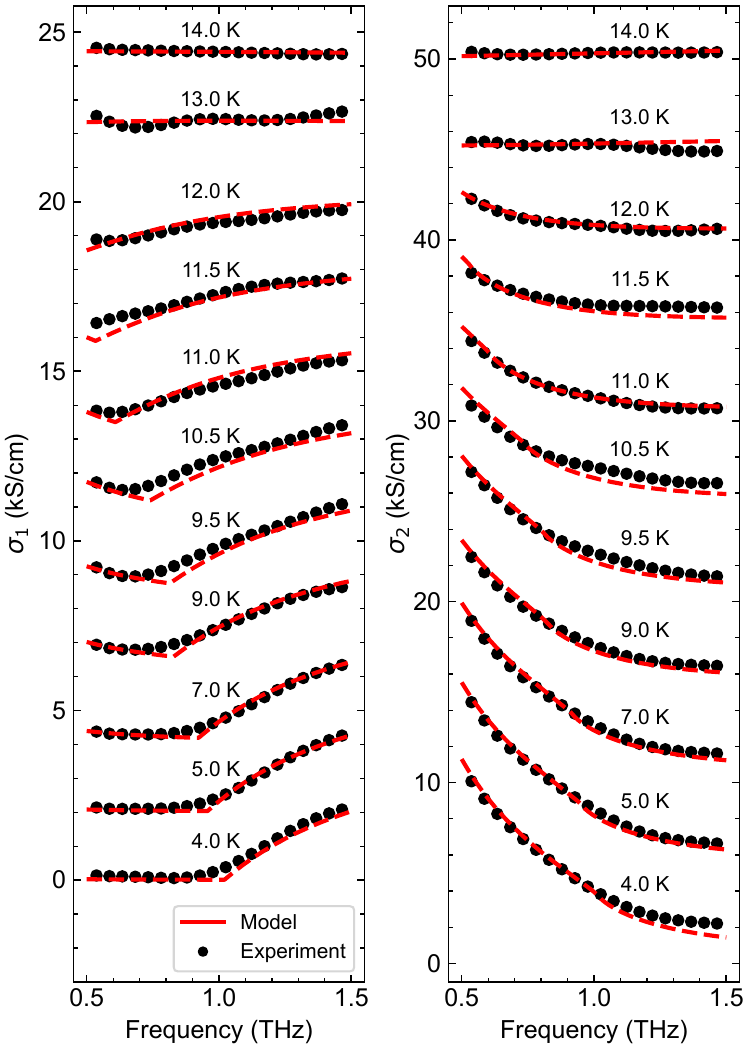}
    \caption{\textbf{The experimental and fitted optical conductivity of NbN film in free space at various temperatures.} Same as Fig.~1d of the main text, but at different temperatures. Each curve of $\sigma_1$ ($\sigma_2$) is vertically offset by 2000\,S/cm (5000\,S/cm) for clarity. 
	}
    \label{fig:Sigma_NbN_free}
\end{figure}

The DC conductivity $\sigma_{\rm{DC}}$ and relaxation time $\tau$ were obtained by fitting the experimental normal-state optical conductivity of the NbN film at $T=14$\,K with the Drude formula
\begin{equation}\label{eq:drude-model}
    \sigma_{\rm{Drude}}(\omega) =\sigma_{1,\rm{Drude}}(\omega) + i \sigma_{2,\rm{Drude}}(\omega)= \frac{\sigma_{\rm{DC}}}{1+\omega^{2}\tau^{2}}+i\frac{\sigma_{\rm{DC}}\omega\tau}{1+\omega^{2}\tau^{2}}.
\end{equation}
The obtained values from the fits were $\sigma_{\rm{DC}}=(4.44\pm0.02)\times10^2$\,S/cm, and $\tau=70\pm3$\,fs. These two values were fixed and used to simulate the optical conductivity for the superconducting states at different temperatures below $T_\mathrm{c}$. 
Figure~\ref{fig:Sigma_NbN_free} shows the experimental and fitted normal-state optical conductivity of the NbN film at $14$\,K. 

\begin{figure}[!htbp]
    \centering
    \includegraphics[width=0.8\linewidth]{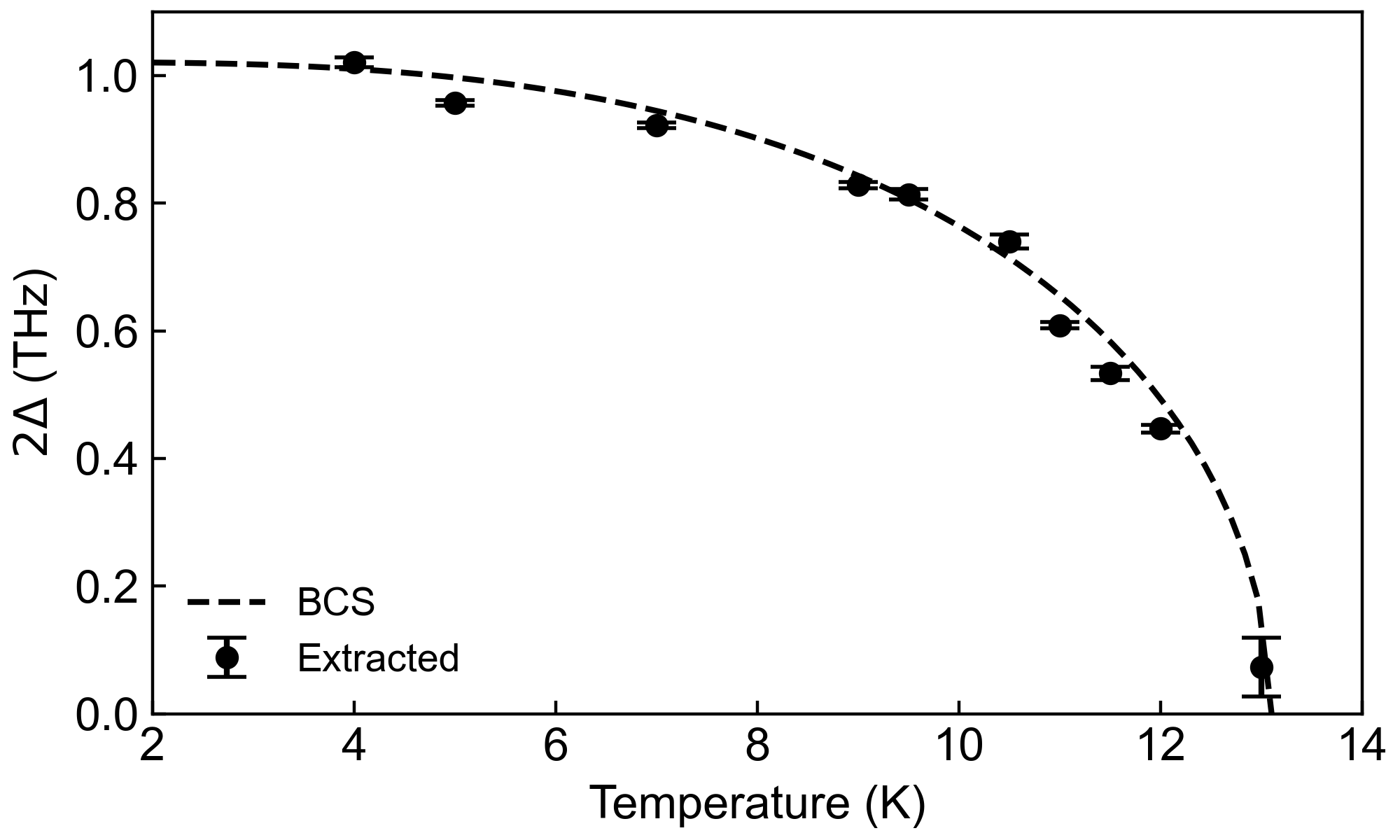}
    \caption{\textbf{Temperature-dependent superconducting optical gap $2\Delta(T)$ for the NbN thin film extracted from the experimental optical conductivity data using the Zimmermann formula [Eq.~\eqref{eq:zimmermann-formula}].} The extracted gap values (black dots) are consistent with the interpolation formula (the dashed line), $\Delta(T) \approx \Delta_{0} \tanh\left(1.74\sqrt{(T_\mathrm{c}/T)-1}\right)$, where $\Delta_{0} = \Delta(T=0)$; see Ref.\,\cite{evtushinsky_momentum-resolved_2009}. The error bars represent fitting uncertainties. } 
    \label{fig:extracted-gaps-NbN}
\end{figure}

To compute the optical conductivity of the NbN film in the superconducting state at a given temperature $T$, we need to obtain the corresponding superconducting gap at that temperature, $\Delta(T)$. 
We extracted $\Delta(T)$ by fitting the experimental optical conductivity of the NbN film at that temperature using the Zimmermann model, with $\sigma_{\rm{DC}}=4.44\times10^2$\,S/cm and $\tau=70$\,fs fixed, as shown in fig.~\ref{fig:Sigma_NbN_free} and fig.~\ref{fig:extracted-gaps-NbN}. 

\section{Experimental and theoretical transmittance of the bare cavity}

\begin{figure}[htbp]
    \centering
    \includegraphics[width=0.85\linewidth]{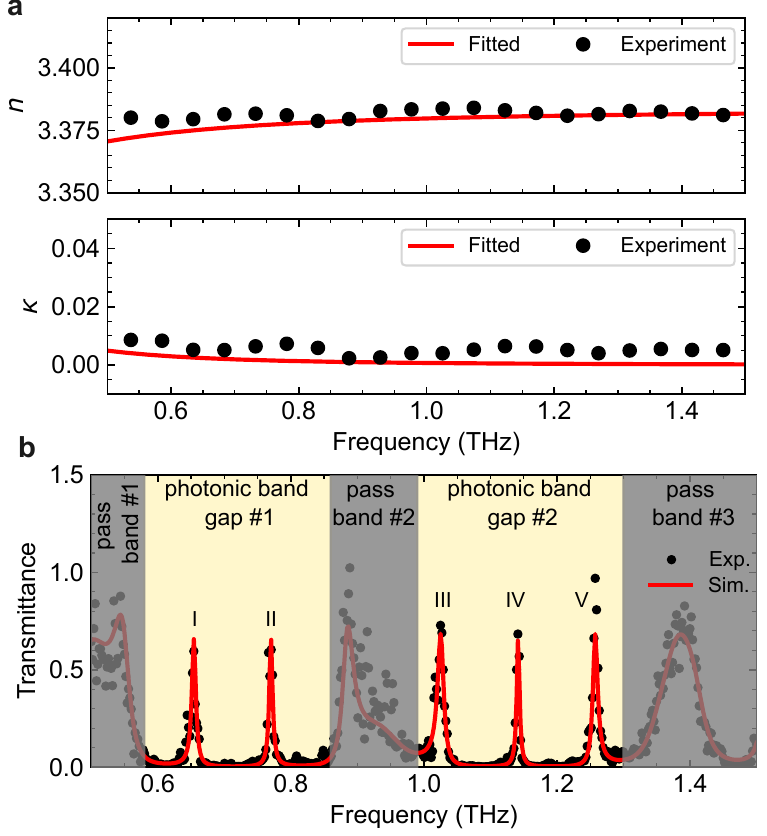}
    \caption{\textbf{The complex refractive index of the thick silicon defect layer and the corresponding bare 1D-PCC transmittance spectra.} (\textbf{a})~shows the complex refractive index $N(\omega)=n(\omega)+i\kappa(\omega)$ of an intrinsic silicon plate (the central defect layer of the 1D-PCC) at $6$\,K. Black dots represent the experimental data, and solid lines represent the fits. (\textbf{b})~ shows the experimental and simulated transmittance using the transfer matrix method of the bare cavity. The complex refractive index we used for the thin silicon mirrors was $3.354+0.0178i$.}
    \label{fig:bare-cavity}
\end{figure}

\begin{figure}[h]
    \centering
    \includegraphics[width=0.8\linewidth]{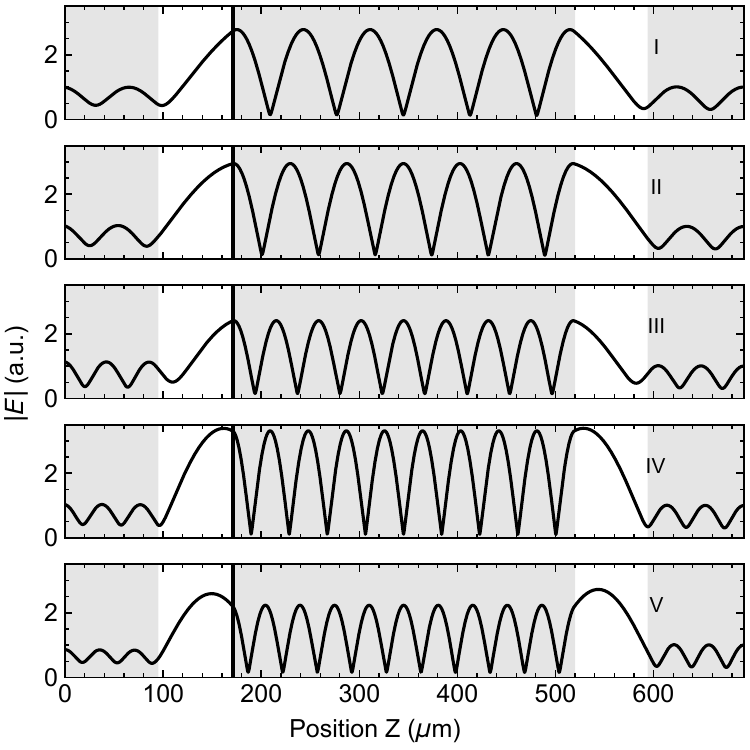}
    \caption{\textbf{The electric field distribution inside the bare 1D-PCC.} We used the TMM method to calculate the electric field distribution inside the bare 1D-PCC and confirmed that the electric field is maximized at the intended NbN film position (vertical black line) at the cavity mode frequencies. 
	}
    \label{fig:Efield_bare}
\end{figure}

Before simulating the transmittance of the superconducting NbN thin film within a cavity, we first analyzed the bare cavity without the NbN film. The structure of the bare cavity is shown in Fig.~1d of the main text.
The frequency-dependent dielectric function of the silicon substrate, $\varepsilon_\mathrm{Si}(\omega)$, can be given in the Drude--Lorentz form,
\begin{equation}
\varepsilon_\mathrm{Si}(\omega) = \varepsilon_{\mathrm{bg}} + \omega_\mathrm{p}^{2}\left(\frac{s_{0}}{-\omega^{2}-i\omega\Gamma_{0}}+\frac{s_{1}}{\omega_{1}^{2}-\omega^{2}-i\omega\Gamma_{1}}\right),
\end{equation}
where $s_0$ ($s_1$) and $\Gamma_0$ ($\Gamma_1$) are the oscillator strength and linewidth of the Drude (Lorentz) term, $\omega_1$ is the resonance frequency of the Lorentzian, $\omega_\mathrm{p}$ is the plasma frequency, and $\varepsilon_{\mathrm{bg}}$ is the background dielectric constant.  The corresponding complex refractive index is $N_\mathrm{Si}(\omega)=\sqrt{\varepsilon_\mathrm{Si}(\omega)} = n_\mathrm{Si}(\omega) + i \kappa_\mathrm{Si}(\omega)$, where $n_\mathrm{Si}(\omega)$ and $\kappa_\mathrm{Si}(\omega)$ denote the refractive index and extinction coefficient, respectively.
For the two thin silicon mirrors, we fixed the complex refractive index at a constant value of $ 3.354 + 0.0178i$, as determined from THz-TDS measurements. 
The model parameters were fitted to match the experimental transmittance of the bare cavity. We found that the Lorentz term did not contribute to the fitted complex refractive index (i.e., $s_1=0$), and we report the Drude parameters as follows: $\varepsilon_{\mathrm{bg}}=11.44\pm0.01$, $\omega_\mathrm{p}/2\pi = 0.020 \pm 0.003\,\mathrm{THz}$, $s_0 = 60 \pm 3$, and $\Gamma_0/2\pi = 0.19 \pm 0.09\,\mathrm{THz}$.

We simulated the THz transmittance using the transfer matrix method (TMM) implemented in the open-source Python package \textit{tmm}~\cite{byrnes_2020}, assuming normal incidence and $s$-polarized light.
The simulated THz spectrum (lower panel of fig.~\ref{fig:bare-cavity}) reproduces the experimental results, and the fitted real and imaginary parts of the refractive index (upper and central panels) agree with the measured complex refractive index of the silicon substrate. In the simulation, we confirmed that the electric field distribution of our 1D-PCC is maximized at the surface of the silicon substrate at the five cavity mode frequencies, as shown in fig.~\ref{fig:Efield_bare}.

We performed Lorentzian fit to the transmittance of the bare 1D-PCC. The results are summarized in Table~\ref{tab:sup_bare_Q}. Note that the $Q$ factor values are on the order of 100, which is similar to the values achieved in our previous work~\cite{zhang_collective_2016,li_vacuum_2018} and higher than those of other types THz cavities typically used in ultrastrong light--matter coupling studies [see, e.g., Ref.~\cite{scalari_ultrastrong_2012,bayer_terahertz_2017}]. 

\begin{table}[h]
\caption{\textbf{Cavity parameters extracted from experimental transmittance of bare 1D-PCC.} The quality factor is given by $Q = \omega_0/\kappa$, where $\omega_0/2\pi$ is the peak center frequency and the photon decay rate, $\kappa/2\pi$, is the full width at half maximum (FWHM) of the peak.}\label{tab:sup_bare_Q}%
\begin{tabular}{ccccc}
\toprule
Mode No.\ & Amplitude & $\omega_0/2\pi$ (THz) & $\kappa/2\pi$ (GHz) & $Q$\\
\midrule
        I & 0.51 & 0.654 & 8.6 & 76\\
		II & 0.61 & 0.769 & 8.4 & 92\\
		III & 0.68 & 1.025 & 13.6 & 75\\
        IV & 0.72 & 1.141 & 5.5 & 208\\
        V & 0.92 & 1.258 & 7.4 & 169\\
\botrule
\end{tabular}
\end{table}

\section{Simulations of the transmission spectrum of the NbN thin film inside the cavity}

\begin{figure}[!t]
    \centering
    \includegraphics[width=0.9\linewidth]{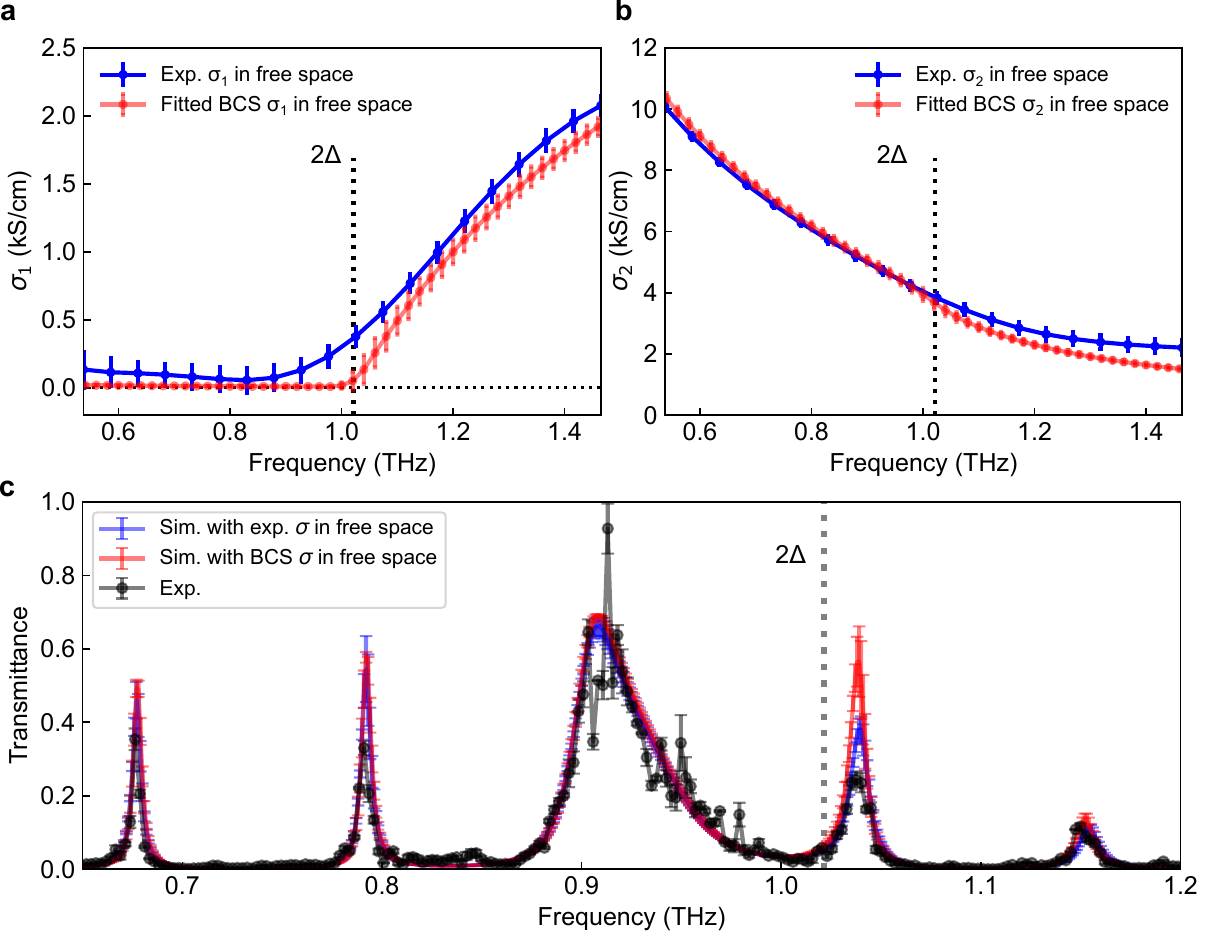}
    \caption{\textbf{Optical conductivity of the NbN thin film in free space and transmittance of the NbN thin film inside the cavity at $4$\,K.} (\textbf{a}) and (\textbf{b})~display the real and imaginary optical conductivity, respectively, from experimental measurements and simulations using the Zimmerman model with the extracted superconducting optical gap, $2\Delta(T)$, at $4$\,K (indicated by the dotted vertical line). (\textbf{c})~compares the experimental transmittance with the simulated one obtained from the experimental and simulated optical conductivity from (\textbf{a}) and (\textbf{b}).}
    \label{fig:opt-cond-spec-NbN-cavity-4K}
\end{figure}

We placed the 10-nm-thick NbN thin film on top of the silicon substrate inside the cavity.
We computed the optical conductivity of the NbN film, $\sigma(\omega,T)$, using the Zimmermann model, 
and then obtained the complex dielectric function through
\begin{equation}
    \varepsilon_\mathrm{r}(\omega) = \varepsilon_{\infty} + i\,\frac{\sigma(\omega)}{\varepsilon_{0}\omega}.
\end{equation}
The complex refractive index was then calculated as $N(\omega)=\sqrt{\varepsilon_\mathrm{r}(\omega)}$.
The resulting simulated real and imaginary parts of the optical conductivity from the Zimmermann model are shown in fig.~\ref{fig:opt-cond-spec-NbN-cavity-4K}A and~\ref{fig:opt-cond-spec-NbN-cavity-4K}B, respectively, compared with the experimental ones.
We then used the experimental and simulated optical conductivity (fig.~\ref{fig:Sigma_NbN_free}) to compute the complex dielectric function, which was fed into the TMM to simulate the corresponding transmittance in the same manner as the bare cavity simulations.
The comparison of the simulated transmittance with the experimental one is shown in fig.~\ref{fig:opt-cond-spec-NbN-cavity-4K}c.
The simulated transmittance using the optical conductivity from the Zimmerman model (red line) agrees with that based on the experimental optical conductivity (blue line), except for the peak just above the superconducting gap, $2\Delta(T=4\ \rm{K})$. 
The difference in that peak results from the discrepancy between the experiment and model. 
However, those simulated peaks below the gaps agree well.
Therefore, we mainly focus on the comparison between the experimental and simulated spectra below the superconducting gaps.
We can see that the intensities of the two simulated peaks, centered around $0.7$ and $0.8$\,THz, overestimate the experimental ones, indicating that the cavity photons modify the optical conductivity of NbN inside the cavity.

\section{First-principles calculations of the superconducting gap and transition temperature of NbN}

The ground state of bulk cubic NbN ($\delta$-NbN) was obtained using the PBEsol exchange-correlation functional and the norm-conserving pseudopotential from PseudoDojo~\cite{van2018pseudodojo}. The calculations were performed in Quantum Espresso (QE)~\cite{QE-2009, Giannozzi_2017}. 
The kinetic energy cutoff was set to 80 Rydberg, and the Monkhorst-Pack $\mathbf{k}$-grid was chosen as $16\times16\times16$ to converge the ground state energy within the error of $1$\,meV/atom.
We used the Marzari-Vanderbilt smearing function with a relatively large smearing value of $0.18$\,Rydberg to avoid the imaginary phonon frequencies~\cite{ivashchenko2010phase}.
The relaxed lattice constant of NbN was $a=4.407\mathring{\mathrm{A}}$, which is close to the experimental value of 4.392$\mathring{\mathrm{A}}$~\cite{brauer1960nitrides}. The convergence threshold for forces on each atom was $0.001$\,Ryd/Bohr.

The phonon properties within the harmonic approximation were calculated using density functional perturbation theory (DFPT)~\cite{baroni2001phonons} in QE. %
We used a $\mathbf{q}$-grid of $8\times 8\times 8$ to obtain the corresponding dynamical matrix elements for each $\mathbf{q}$ point, which were Fourier-transformed to obtain the interatomic force constants. 
The interatomic force constants were then used to interpolate the harmonic phonon dispersion for an arbitrary $\mathbf{q}$ point.
The anharmonic phonons were obtained by the anharmonic special displacement method (A-SDM)~\cite{zacharias2023anharmonic} implemented in EPW~\cite{lee2023electron}. 
We used a $2\times2\times2$ supercell to compute the force constants for the anharmonic phonons. 
By the Fourier transform of the anharmonic force constants, we obtained the dynamical matrices for an $8\times 8\times 8$ $\mathbf{q}$-grid, and then Fourier-transformed the resulting dynamical matrices to get the anharmonic interatomic force constants for an $8\times8\times8$ grid in real space, which were used for subsequent electron-phonon coupling calculations. 

To reduce the computational cost of electron-phonon coupling calculations, we used the Wannier interpolation technique~\cite{giustino_electron-phonon_2017}.
We used a coarse $8\times8\times8$ $\mathbf{k}$-grid to construct the Wannier functions in Wannier90~\cite{Pizzi2020}, for which the initial projections were chosen as the 2$p$ orbitals for nitrogen and 4$d$ orbitals for niobium.
We computed the electron--phonon couplings in the Bloch basis on the coarse $\mathbf{k}$-grid and $\mathbf{q}$-grid using EPW~\cite{lee2023electron,margine2013anisotropic}, and Fourier-transformed them into the Wannier basis.
With electron--phonon couplings in the Wannier basis, we were able to obtain any electron-phonon coupling at arbitrary $\mathbf{k}$ and $\mathbf{q}$ points via the inverse Fourier transformation.
To obtain the superconducting gap, we solved the anisotropic Eliashberg equations using EPW with a fine $\mathbf{k}$-grid of $72\times72\times72$ points and a fine $\mathbf{q}$-grid of $36\times36\times36$. 
The Coulomb pseudopotential is set to $\mu^*=0.14$.

To include electron--photon interactions from cavity photons, we used quantum electrodynamical density-functional theory, which included an additional electron--photon exchange potential in the Kohn-Sham Hamiltonian~\cite{lu_cavity-enhanced_2024}. 
Our simulations utilized two cavity modes with polarization along the $x$ and $y$ directions. The ratio of the mode strength $\lambda_{\alpha}$ to the bare frequency $\omega_\alpha$ for the two modes was chosen as $\lambda_\alpha/\omega_\alpha=0.1$.
We fixed the lattice constant and relaxed the atomic positions in the unit cell, but the atomic positions remained the same as in the outside cavity case.
Other computational conditions were the same as those outside the cavity.

\begin{figure}[htbp]
    \centering
    \includegraphics[width=\textwidth]{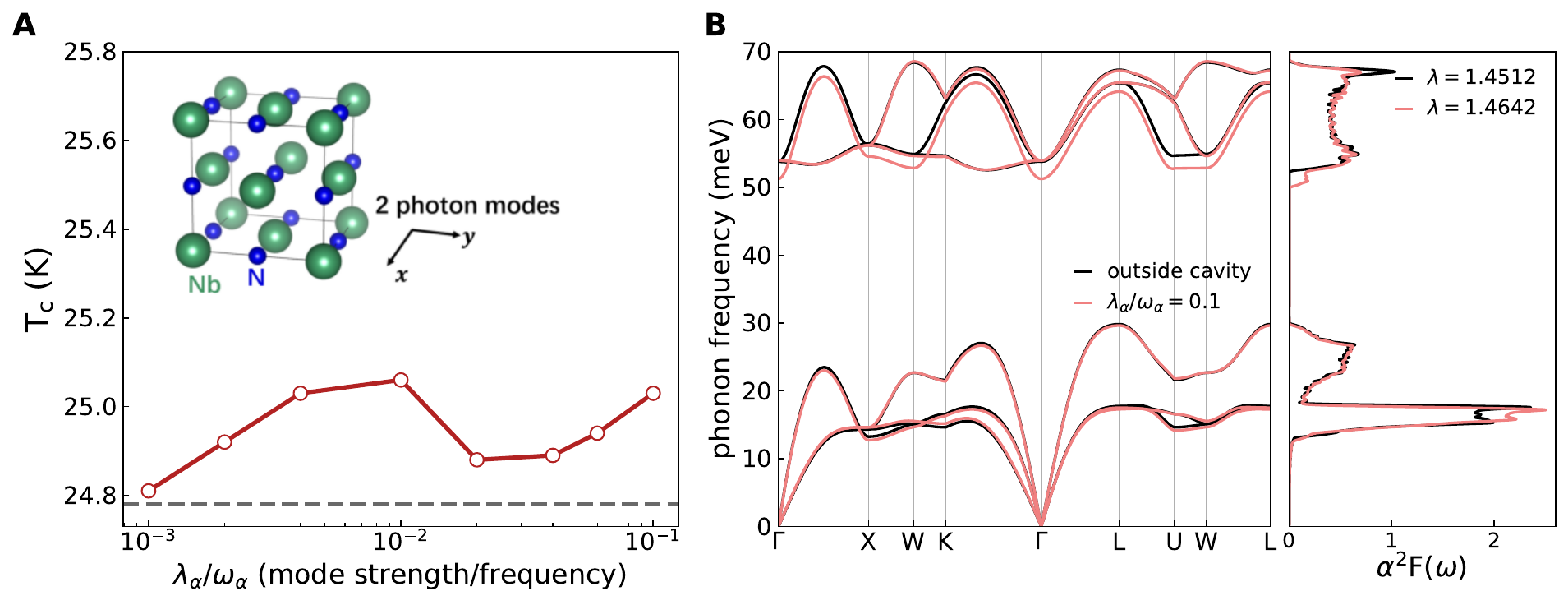}
    \caption{\textbf{First principles calculations of NbN.}
    (\textbf{A})~The Allen--Dynes superconducting transition temperature of NbN as a function of the ratio of the mode strength $\lambda_\alpha$ to the bare photon frequency $\omega_\alpha$. 
    (\textbf{B})~Cavity-modified phonon dispersion and Eliashberg function. $\lambda$ is the total electron--phonon coupling. }
    \label{fig:NbN-allen-dynes} 
\end{figure}

To explore how the cavity modes affect the superconducting transition temperature $T_{\mathrm{c}}$ of NbN, we first used a computationally cheap way, the Allen-Dynes formula~\cite{allen1975transition}, 
\begin{equation}
    T_{\mathrm{c}}^{\rm{AD}}=\frac{\omega_{\rm{log}}}{1.2}\exp{\left[\frac{-1.04(1+\lambda)}{\lambda-\mu^*(1+0.62\lambda)}\right]},
\end{equation}
where $\lambda$ is the total electron--phonon coupling, $\mu^*$ is the Coulomb pseudopotential, and
\begin{equation}
    \omega_{\rm{log}}=\exp{\left[\frac{2}{\lambda}\int_0^{+\infty}\alpha^2F(\omega)\log{(\omega)}/\omega d\omega\right]},
\end{equation}
with $\alpha^2F(\omega)$ being the Eliashberg function and $\omega$ being the phonon frequency.
Figure~\ref{fig:NbN-allen-dynes}(a) shows the simulated $T_{\mathrm{c}}^{\rm{AD}}$ of NbN as a function of the ratio of the mode strength to the bare frequency. 
For NbN in free space, we obtained $T_{\mathrm{c}}^{\rm{AD}}=24.78$\,K. Overall, $T_{\mathrm{c}}^{\rm{AD}}$ does not change too much within $0.3$\,K even in a large value of $\lambda_{\alpha}/\omega_{\alpha} =0.1$, which is the upper limit of a realistic device~\cite{lu_cavity-enhanced_2024}. 
The non-monotonic behavior of $T_{\mathrm{c}}^{\rm{AD}}$ is from the non-monotonic behavior of $\lambda$, which is due to the redistribution of the phonon modes around $15$\,meV. Figure~\ref{fig:NbN-allen-dynes}(B) shows the cavity-modified phonon dispersion and Eliashberg function for $\lambda_\alpha/\omega_\alpha=0.1$. Therefore, we do not expect any observable change of $T_{\mathrm{c}}$ in the experiment, as the actual $\lambda_{\alpha}/\omega_{\alpha}$ is typically much lower than 0.1. 

\begin{figure}[htbp]
    \centering
    \includegraphics[width=0.65\textwidth]
    {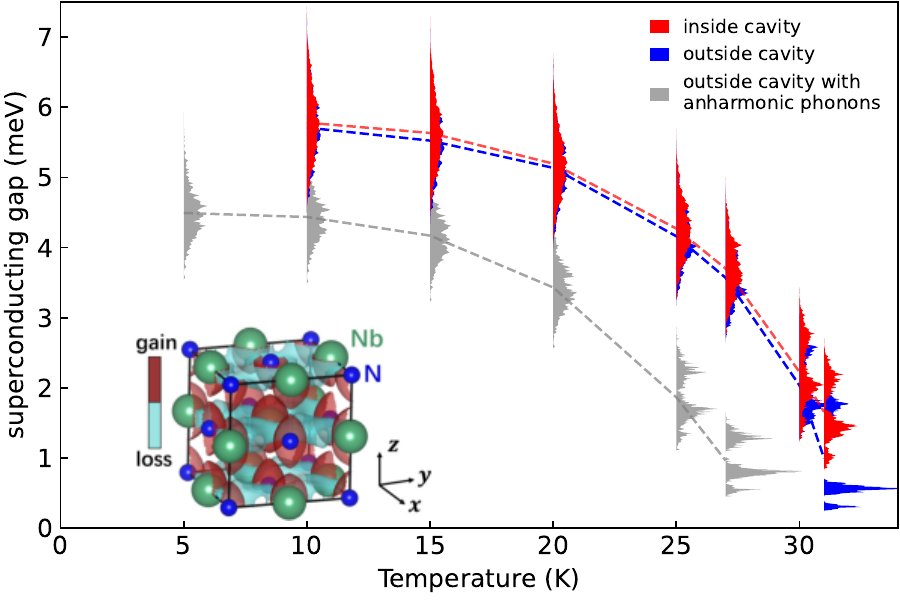}
    \caption{\textbf{Superconducting gap for NbN inside and outside cavity.} In the lower left of the figure, the difference of the electron density between the inside and outside of the cavity is shown. For the cavity case, we chose the electron--photon coupling strength $\lambda_\alpha/\omega_\alpha=0.1$.}
    \label{fig:NbN-superconducting-gaps}
\end{figure}

Next, we used another more accurate but computationally expensive approach, anisotropic Eliashberg equations, to estimate how the superconducting gaps and $T_{\mathrm{c}}$ are modified for NbN inside a cavity. 
Figure~\ref{fig:NbN-superconducting-gaps} shows the temperature-dependent superconducting gap obtained from solving the Eliashberg equations. 
The $T_{\mathrm{c}}$ can be estimated at the temperature where the superconducting gap vanishes.
The $T_{\mathrm{c}}$'s are about $32$\,K for both cases where the NbN is outside and inside a cavity with the ratio of $\lambda_\alpha/\omega_\alpha=0.1$. 
The superconducting gap for NbN is computed down to $10$\,K. 
Our computed $T_{\mathrm{c}}$ was overestimated, compared to the experimental one, $T_{\mathrm{c}}=13.1$K.
The discrepancy between the computational and experimental values may be due to impurity effects~\cite{kalal2021effect}, phonon anharmonicity~\cite{setty2024anharmonic}, or non-adiabatic effects~\cite{calandra2010adiabatic}, which were neglected in the calculations.
When the anharmonic phonons are considered, $T_\mathrm{c}$ reduces to $27$\,K for the case without the cavity. 
We focus more on whether the cavity modifies the superconducting gap, rather than on the absolute value of $T_{\mathrm{c}}=13.1$\,K. 
The harmonic phonon results are sufficient to demonstrate that the change in the superconducting gap inside the cavity is only slight compared to the case without a cavity at low temperatures. 
When NbN is placed inside a cavity with two photon modes polarized along the $x$ and $y$ directions, the electron density of the \textit{normal phase} decreases in the $xy$-plane and accumulates along the $z$ direction; see the inset of fig.~\ref{fig:NbN-superconducting-gaps}.
These results indicate that the cavity modifies the electron density, even though the superconducting gap shows only minimal changes.

\section{Temperature dependence of the cavity modes parameter of the NbN-cavity system}

\begin{figure}[!hb]
    \centering
    \includegraphics[width=\linewidth]{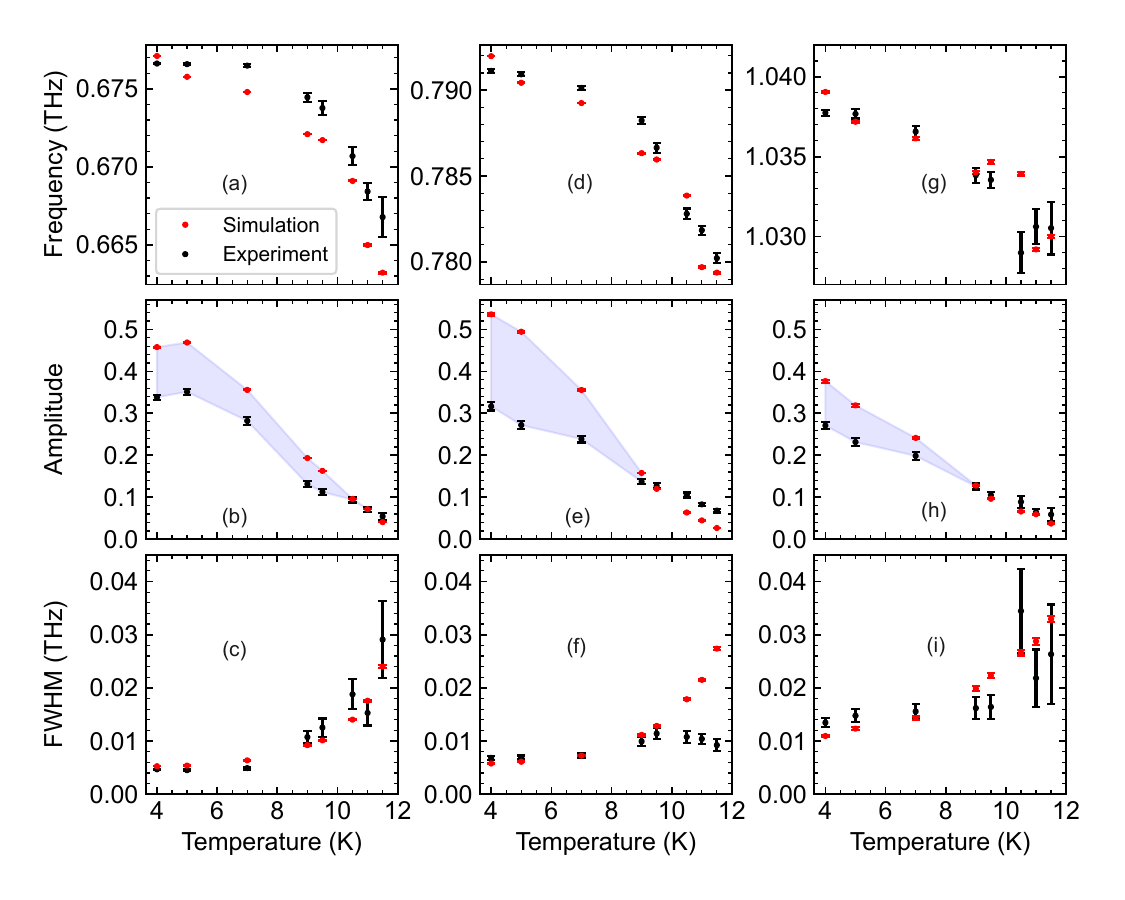}
    \caption{\textbf{Results of the Lorentzian fitting of peaks I, II, and III at various temperatures in the superconducting state of NbN.} The shaded regions highlight the amplitude differences between the TMM simulation (red) and the experiment (black). The error bars represent the uncertainties associated with the Lorentzian fitting procedure. Note that the peak amplitude decreases at high temperature, hence the larger fitting error bars. (\textbf{a-c}) Peak I. (\textbf{d-f}) Peak II. (\textbf{g-i}) Peak III.} 
    \label{Peak_analysis}
\end{figure}

The results of the Lorentzian fitting of peaks I, II, and III at various temperatures in the superconducting state of NbN are shown in fig.~\ref{Peak_analysis}. Note that the peak amplitude decreases at high temperature, hence the larger fitting error bars, which represent the uncertainties associated with the Lorentzian fitting procedure. When comparing peak frequencies (fig.~\ref{Peak_analysis}(a,d,f)) and linewidths (fig.~\ref{Peak_analysis}(c,f,i)) between the simulation (red) and the experiment (black), the differences are minor. However, the peak amplitudes show the most prominent changes, which are indicated by the shaded regions in fig.~\ref{Peak_analysis}(b,e,h). Additionally, it can be observed that the amplitude differences set in at different temperatures for the three peaks, as related to the change in the optical gap with temperature. 

\section{The extracted in-cavity optical conductivity of the NbN film at various temperatures}

In this section, we report the extracted in-cavity optical conductivity of the NbN film at various temperatures, except for $T=4\,\mathrm{K}$, for which see Fig.~3a of the main text. As the temperature increases, the below-gap discrepancy in $\sigma_1$ vanishes. This is consistent with the transmittance spectra reported in Fig.~2a of the main text, as the transmittance discrepancy also vanishes at higher temperatures. On the other hand, the in-cavity imaginary part $\sigma_2$ does not show significant differences compared to the free-space counterpart. Note that the fitting became less reliable at $T>9\,\mathrm{K}$ as the cavity peaks amplitude decreased. 
\begin{figure}[hbtp]
    \centering
    \includegraphics[width=0.85\linewidth]{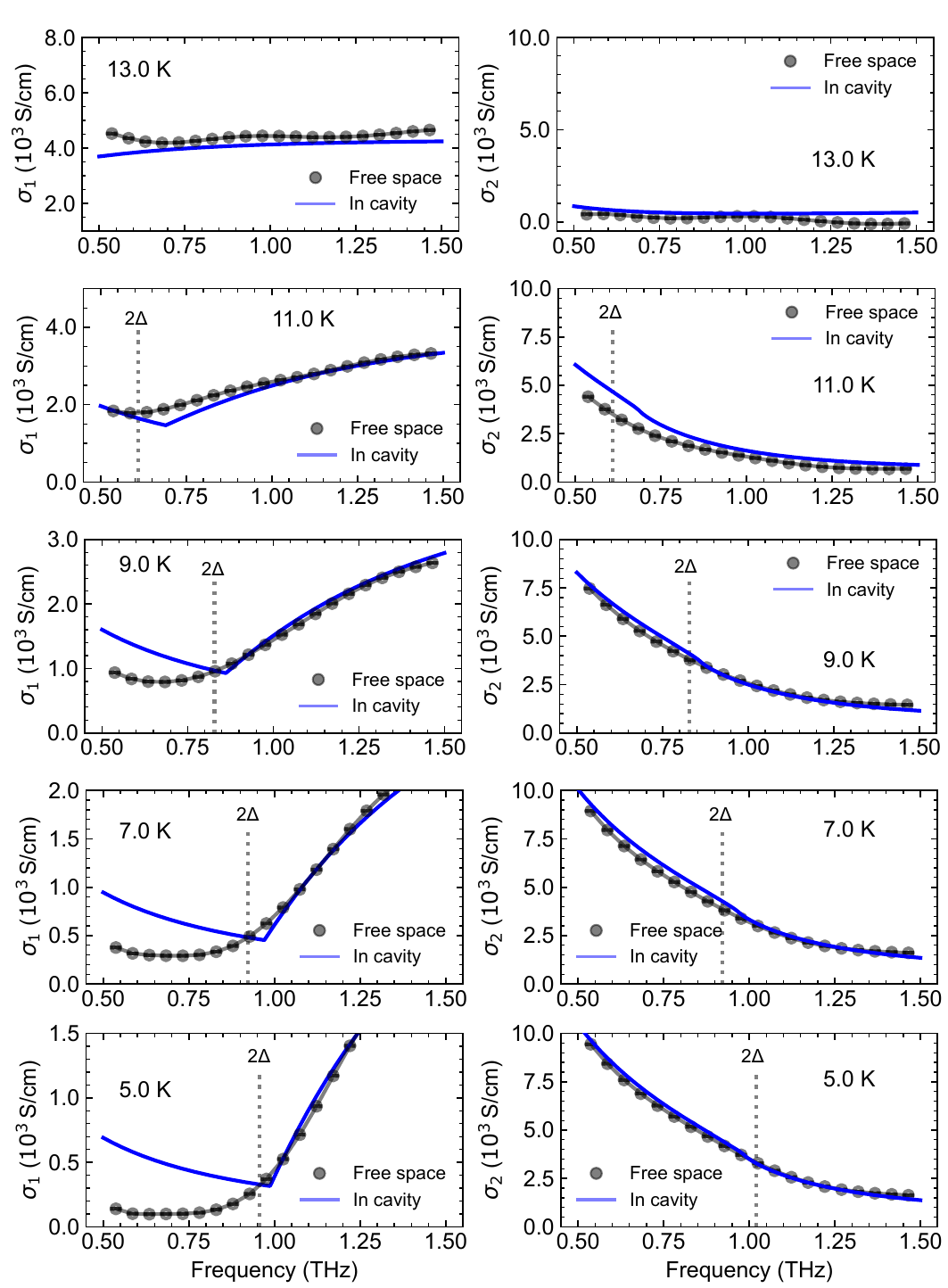}
    \caption{\textbf{In-cavity optical conductivity of the NbN film at various temperatures.} Significant modifications below the gap have been observed in in-cavity $\sigma_1$ below $9\,$K, which gradually weakens as the temperature approaches $T_\mathrm{c}$. } 
    \label{fig:Mod_sigma_Temp}
\end{figure}

\end{document}